\documentclass[12pt,preprint,nofootinbib,superscriptaddress,prd]%
{revtex4}

\long\def\dump#1{}

\def\beq{\begin{equation}}
\def\eeq{\end{equation}}
\usepackage[dvips]{graphicx}
\usepackage{psfig}
\usepackage{epsfig}
\usepackage{color}

\def\iso#1#2{\mbox{${}^{#2}{\rm #1}$}}
\def\k4#1{\iso{K}{4#1}}
\def\u23#1{\iso{U}{23#1}}
\def\th23#1{\iso{Th}{23#1}}

\begin{document}

\title {Probing the Earth's interior with a large-volume
liquid scintillator detector}

\author{Kathrin A.~Hochmuth}
\email{hochmuth@ph.tum.de}
\affiliation{Technische Universit\"at M\"unchen,
Physik Department E15, James-Franck-Strasse, 85748 Garching, Germany}
\affiliation{Max-Planck-Institut f\"ur Physik
(Werner-Heisenberg-Institut), F\"ohringer Ring 6, 80805 M\"unchen,
Germany}

\author{Franz~v.~Feilitzsch}
\affiliation{Technische Universit\"at M\"unchen,
Physik Department E15, James-Franck-Strasse, 85748 Garching, Germany}

\author{Brian~D.~Fields}
\affiliation{Center for Theoretical Astrophysics,
Department of Astronomy, University of Illinois,
Urbana, IL~61801, USA}

\author{Teresa~Marrod\'an~Undagoitia}
\affiliation{Technische Universit\"at M\"unchen,
Physik Department E15, James-Franck-Strasse, 85748 Garching, Germany}

\author{Lothar~Oberauer}
\affiliation{Technische Universit\"at M\"unchen,
Physik Department E15, James-Franck-Strasse, 85748 Garching, Germany}

\author{Walter~Potzel}
\affiliation{Technische Universit\"at M\"unchen,
Physik Department E15, James-Franck-Strasse, 85748 Garching, Germany}

\author{Georg~G.~Raffelt}
\affiliation{Max-Planck-Institut f\"ur Physik
(Werner-Heisenberg-Institut), F\"ohringer Ring 6, 80805 M\"unchen,
Germany}

\author{Michael~Wurm}
\affiliation{Technische Universit\"at M\"unchen,
Physik Department E15, James-Franck-Strasse, 85748 Garching, Germany}

\begin{abstract}
A future large-volume liquid scintillator detector would provide a
high-statistics measurement of terrestrial antineutrinos originating
from $\beta$-decays of the uranium and thorium chains.  In addition,
the forward displacement of the neutron in the detection reaction
$\bar\nu_e+p\to n+e^+$ provides directional information.  We
investigate the requirements on such detectors to distinguish
between certain geophysical models on the basis of the angular
dependence of the geoneutrino flux.  Our analysis is based on a
Monte-Carlo simulation with different levels of light yield,
considering both unloaded and gadolinium-loaded scintillators. We
find that a 50~kt detector such as the proposed LENA (Low Energy
Neutrino Astronomy) will detect deviations from isotropy of the
geoneutrino flux significantly.  However, with an unloaded
scintillator the time needed for a useful discrimination between
different geophysical models is too large if one uses the
directional information alone. A Gd-loaded scintillator improves the
situation considerably, although a 50~kt detector would still need
several decades to distinguish between a geophysical reference model
and one with a large neutrino source in the Earth's core. However, a
high-statistics measurement of the total geoneutrino flux and its
spectrum still provides an extremely useful glance at the Earth's
interior.
\end{abstract}

\maketitle

\section{Introduction}

The first measurement of the $\bar\nu_e$ flux from natural radioactive
elements in the Earth~\cite{Araki:2005qa} has triggered a lot of
excitement about the future of ``neutrino geology''
\cite{Fiorentini:2005mr, Fiorentini:2005cu, geonu, deMeijer:2004wq}.
The geoneutrino\footnote{It is understood that this term refers to
electron-antineutrinos emitted by the Earth's natural radioactive
elements.} flux could deliver new information about the interior of
the Earth, in particular its radiochemical composition, and thus shed
new light on the question of how the Earth and other planets formed.
Such an ambitious programme requires detectors of the next generation
that are able to provide much larger event rates.

One possible future detector that may well serve this purpose is LENA
(Low Energy Neutrino Astronomy) that has been proposed by several of
us~\cite{Oberauer:2005kw} for high-statistics solar neutrino
spectroscopy, for spectroscopy of the cosmic diffuse supernova
neutrino background (DSNB), as a detector for the next galactic
supernova, and to search for proton
decay~\cite{Undagoitia:12uu}. Present design studies for LENA assume
50~kt of liquid scintillator that would provide a geoneutrino rate of
roughly one thousand events per year from the dominant
\begin{equation}\label{eq:detect}
\bar\nu_e+p\to n+e^+
\end{equation}
inverse beta-decay reaction.

While liquid scintillator detectors do not provide direct angular
information, indirectly one can retrieve directional information
because the final-state neutron is displaced in the forward
direction. The offset between the $e^+$ and the neutron-capture
locations can be reconstructed, although with large uncertainties.
The CHOOZ reactor neutrino experiment has demonstrated a nontrivial
directional sensitivity in that it was able to locate the reactor
within $18^\circ$ half-cone aperture (68\% C.L.) on the basis of 2500
reconstructed events in a Gd-loaded
scintillator~\cite{chooz}. Therefore, it is natural to study the
requirements for a future large-volume liquid scintillator detector to
discriminate between different geophysical models of the Earth that
differ both by their total neutrino fluxes and the neutrino angular
distributions.

We will consider two types of detectors. Motivated by current design
studies for LENA we will consider a 50~kt detector using a PXE based
scintillator. However, in this case it is difficult to locate the
neutron-capture event on protons because a single 2.2~MeV gamma is
released that travels on average 22.4~cm before its first Compton
interaction. Therefore, the event reconstruction is relatively
poor. As a second case we study a Gd-loaded scintillator where the
neutron capture releases on average 3~photons with a total of 8~MeV,
allowing for a much better event reconstruction.  Moreover, the
spatial resolution can likely be improved beyond the original CHOOZ
experiment if one uses photomultipliers that are fast enough (time
jitter $\sim1$~ns) to use the photon arrival times for the event
reconstruction.

For the geoneutrino flux we will consider a continental and an oceanic
location. In each case we will use a reference model and exotic cases
with an additional strong neutrino source in the Earth's core or with
enhanced neutrino emission from the mantle.

We will begin in Sec.~\ref{sec:LENA} with a discussion of the
principle of geoneutrino detection in large-volume scintillator
detectors as well as possible scintillator properties.  In
Sec.~\ref{sec:geomodels} we introduce our geophysical models.  In
Sec.~\ref{sec:montecarlo} we turn to the main part of our work, a
Monte-Carlo study of the discriminating power of different detectors
between different geophysical models and conclude in
Sec.~\ref{sec:conclusions}.

\section{Geoneutrino detection}
\label{sec:LENA}

\subsection{Directional information from neutron displacement}

In a scintillator detector, geoneutrinos are measured by the inverse
beta-decay reaction Eq.~(\ref{eq:detect}) with an energy threshold of
1.8~MeV. The cross section is
\begin{equation}
\sigma=9.52 \times 10^{-44}~{\rm cm}^2\,\,
\frac{E_+}{\rm MeV} \frac{p_+}{\rm MeV}
\end{equation}
where $E_+$ is the total energy of the positron and $p_+$ its
momentum.  The visible energy $E_{\rm vis}=E_++m_e$ always exceeds
1~MeV because the positron annihilates with an electron of the
target. By measuring the visible energy one can determine the neutrino
energy as $E_\nu \approx E_{\rm vis} + 0.8~{\rm MeV}$ because the
kinetic energy of the neutron is typically around 10~keV and thus
negligible.  After thermalization the neutron is captured by a
nucleus, thus tagging the inverse beta decay reaction.

Kinematics implies that the neutron is scattered roughly in the
forward direction with respect to the incoming neutrino~\cite{vogel},
this being the key ingredient for obtaining directional information.
The maximum scattering angle is
\begin{equation}
\cos\theta_{\rm max}=\frac{\sqrt{2E_\nu \Delta -
(\Delta^2-m_e^2)}}{E_\nu}\,,
\end{equation}
where $\Delta=m_n-m_p$ and $m_n$, $m_p$ and $m_e$ are the masses of
the neutron, proton and positron, respectively. In this extreme
case, the neutron and positron momenta are perpendicular to each
other.  For the maximum relevant geoneutrino energy of 3.2~MeV one
obtains $\cos\theta_{\rm max}=0.79$, which is equivalent to $\theta_{\rm max}=37.8^\circ$. However, most geoneutrinos have
energies much closer to threshold and the maximum angles are much
closer to the forward direction.  The average displacement between
the neutron and positron events is then theoretically found to be
about 1.7~cm~\cite{vogel}.

The reactor experiment CHOOZ, using a Gd-loaded scintillator, has
measured an average neutron displacement from the $e^+$ event of
$1.9 \pm 0.4$~cm~\cite{chooz}. However, once the neutron has been
thermalized by collisions with protons, it diffuses some distance
before being captured so that the actual displacement varies by a
large amount for individual events. In a PXE based scintillator the
average time interval until capture on a proton is 180~$\mu$s,
leading to an uncertainty $\sigma$ of the displacement of about 4~cm
for the x-, y- and z-direction~\cite{vogel}. With Gd loading
$\sigma$ is reduced to approximately 2.4~cm~\cite{vogel} because the
neutron diffusion time is much shorter, on average about
30~$\mu$s~\cite{chooz}.

\subsection{PXE-based scintillator}

One option for the proposed LENA detector is to use a scintillator
based on PXE (phenyl-o-xylylethane, C$_{16}$H$_{18}$). PXE has a high
light yield, it is non hazardous, has a relatively high flashpoint of
145$^\circ$C, and a density of 0.985 g/cm$^3$ \cite{Back:2004zn}. A
possible admixture of dodecane (C$_{12}$H$_{26}$) increases the number
of free protons and improves the optical properties.  A blend of
20\%~PXE and 80\%~dodecane shows a decrease in light yield of about
20\% relative to pure PXE, an attenuation length of about 11~m and an
increase in the number of free protons by~25\%~\cite{wurm}.

In this paper we consider a detector with a total volume of about
$70 \times 10^3$~m$^3$. This could be realized with a cylindrical
detector of 100~m length and 30~m diameter.  An outer water
Cherenkov detector with a width of 2~m acts as a muon veto. In order
to shield against external gamma and neutron radiation a fiducial
volume of about $42 \times 10^3~{\rm m}^3$ with a total number of
$2.5 \times 10^{33}$ free protons as target can be realized using a
scintillator mixture as mentioned above with 20\%~PXE and
80\%~dodecane. In Monte-Carlo calculations the light yield of events
in LENA has been determined~\cite{Undagoitia:12uu}. For events in
the central detector region the yield $N_{\rm pe}$, measured in
photo-electrons (pe) per MeV energy deposition, can be expressed as
$N_{\rm pe} \approx 400~{\rm pe/MeV} \times c$, where $c$ is the
optical coverage which depends on the number and aperture of the
photomultiplier tubes (PMTs).  A maximal coverage $c_{\rm max}
\approx 0.75$ can not be exceeded so that we assume the maximal
light yield to be around 300~pe/MeV.  For instance, the use of
12,000 PMTs with a diameter in aperture of 50~cm would result in an
optical coverage of about 30\% and a light yield $N_{\rm pe} \simeq
120$~pe/MeV.  This can be obtained either by using PMTs like in the
Super-Kamiokande experiment or by smaller PMTs equipped with light
concentrators as they were developed for the Counting Test Facility
(CTF) at the Gran Sasso underground laboratory~\cite{conc}.  For
events off the axis of the cylinder the light yield would be
enhanced.  Hence, low-energy spectroscopy even in the sub-MeV region
should be possible in LENA.

For a detection of the positron-neutron displacement the ability of
the detector to locate the absorption position of both particles is
crucial. The experimental reconstruction of both events is possible
by analyzing the arrival times and the number of photons in each
individual PMT. The position uncertainty depends on the total yield
of registered photo-electrons. In the CTF, the measured position
uncertainty was around 10~cm in each direction for events with 300
photo-electrons and it was shown that the uncertainty scales with
the inverse square-root of that number~\cite{ctf}.  Therefore, we
will use a Gaussian distribution for the uncertainty of the positron
event reconstruction with equal width in each direction~of
\begin{equation}\label{eq:sigma_positron}
\sigma_{e^+}=10~{\rm cm}~\left(\frac{300~{\rm pe/MeV}}{N_{\rm pe}}\,\,
\frac{1~{\rm MeV}}{E_{\rm vis}}\right)^{1/2}
\end{equation}
where $N_{\rm pe}$ is the light yield and $E_{\rm vis}$ the visible
energy released by the positron.

\begin{figure}
\begin{center}
\includegraphics[width=0.6\textwidth]{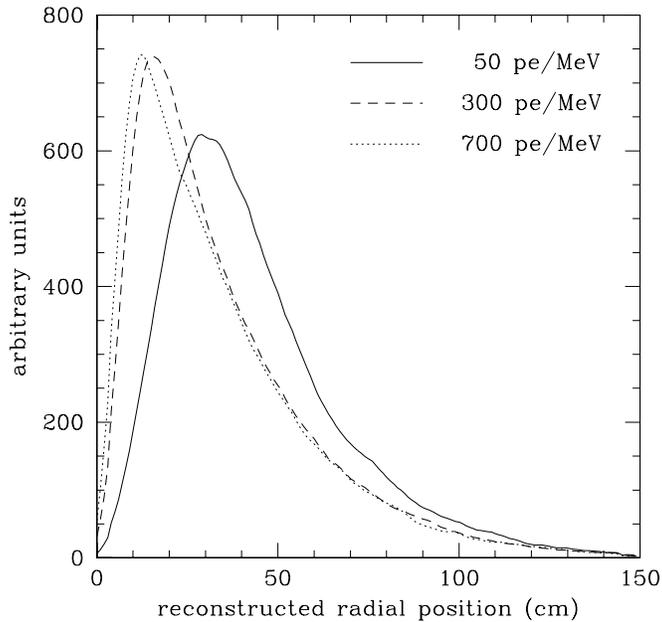}
\caption{\label{fig:1} Monte-Carlo simulation of the radial
distribution of a 2.2~MeV $\gamma$-quantum in an unloaded PXE
scintillator. The curves are for light yields of $N_{\rm pe}=50$,
300 and 700~pe/MeV as indicated.}
\end{center}
\end{figure}

In PXE-based scintillators the neutron is captured by a proton with
nearly 100\% efficiency within an average time interval of about
180~$\mu$s, subsequently emitting a 2.2~MeV gamma.  This photon has a
mean free path of 22.4~cm before its first Compton scattering so that
the event reconstruction is much more uncertain than for the positron
event. We have simulated this case by taking into account multiple
Compton scatterings of the 2.2~MeV gamma.  The position of each gamma
emission, representing the position of the neutron capture, is
reconstructed by composing the energy-weighted sum of each Compton
scattering event, taking into account the instrumental resolution.
The distribution of the reconstructed position in each direction
follows roughly a Lorentzian form. In Fig.~\ref{fig:1} we show the
radial distribution of the reconstructed positions of these events for
light yields of $N_{\rm pe}=50$, 300 and 700~pe/MeV. Increasing the
light yield does not significantly narrow the distribution because its
width is dominated by the large Compton mean free path of the 2.2~MeV
photon.  With reduced light yield the position of the maximum as well
as the mean value of the distribution shifts towards larger
values. This is caused by the increased uncertainty of the
instrumental resolution.

\subsection{Gadolinium-loaded scintillator}

In PXE-based scintillators the neutron is captured by a proton with
nearly 100\% efficiency within an average time interval of about
180~$\mu$s, subsequently emitting a 2.2~MeV gamma. However, one can
also put an additive into the scintillator to enhance the
delayed-neutron signal. Both lithium~\cite{bugey} and
gadolinium~\cite{chooz} loaded scintillators have been used in this
way for neutrino experiments or are planned in the
future~\cite{Ardellier:2004ui}. With a concentration of about 0.1\% by
mass, Gd captures neutrons with 90\% efficiency and isotropically
emits a total energy of about 8~MeV in a gamma cascade with an average
of 3~photons. Note however, that in a Gd-loaded PXE/dodecane scintillator the light yield is reduced by typically 10--20\% compared to an unloaded scintillator.

For Gd-loaded scintillators we use the measurements performed in the
CHOOZ experiment. The neutron response of the detector was measured
with a $^{252}$Cf source and the position uncertainty of a neutron
capture event was 19~cm at $1\sigma$ in each direction~\cite{chooz19}.
The distribution was Gaussian and the light yield was measured to be
$N_{\rm pe}=(125 \pm 5)$~pe/MeV. The reconstruction was performed
using only the information about the amplitude distribution of the
PMTs. Based on these numbers we assume the position resolution of the
neutron event for the LENA detector to be Gaussian with a width in
each direction of
\begin{equation}\label{eq:sigma_neutron}
\sigma_n=19~{\rm cm}\,\,
\left(\frac{125~{\rm pe/MeV}}{N_{\rm pe}}\right)^{1/2}\,.
\end{equation}
In CHOOZ no information of the arrival time was used for the
position reconstruction. However, the CTF measurements demonstrated
that this yields the most valuable information, provided the time
response of the PMTs is fast enough, i.e.\ a time jitter not much
larger than 1~ns. Therefore, in a Gd-loaded scintillator one
probably could achieve a much better neutron-event reconstruction so
that our estimates are conservative.

\subsection{Backgrounds}

KamLAND has reported 152 events in the energy region relevant for
geoneutrinos within a measuring time of 749 days and $3.5 \times
10^{31}$ target protons.  From these events $127 \pm 13$ are due to
background~\cite{kamland1}. The most relevant background for the
KamLAND site are reactor antineutrinos ($80.4 \pm 7.2$ events). For
the LENA detector positioned in the underground laboratory CUPP (Centre for Underground Physics in Pyh\"asalmi) in
 Finland (longitude: 26$^\circ$ 2.709' E,latitude: 63$^\circ$ 39.579' N,
1450 m of rock (4060 m.w.e.)) this background would be reduced by a factor $\simeq$~12, as
the site is far away from reactors.  Hence we expect for LENA at CUPP
a reactor background rate of about 240 events per year in the relevant
energy window from 1.8~MeV to 3.2~MeV.  This background can be
subtracted statistically using the information on the entire reactor
neutrino spectrum up to $\simeq$~8 MeV.

Another important background for KamLAND is induced by radio
impurities. A large concentration of the long-lived isotope
$^{210}$Pb is present in the KamLAND scintillator. In the decay
chain of $^{210}$Pb the $\alpha$-emitting isotope $^{210}$Po is present.
Thus the reaction $^{13}$C$(\alpha,n)^{16}$O can occur, mimicking
the signature of geoneutrinos due to neutron scattering on protons
and the subsequent neutron capture. The number of these background
events in KamLAND is estimated to be $42\pm 11$~\cite{kamland1}.
However, with an enhanced radiopurity of the scintillator, the
background can be significantly reduced. Taking the radio purity
levels of the CTF detector, where a $^{210}$Po activity of
$35\pm12/\rm{m^3d}$ in PXE has been observed~\cite{Back:2004zn},
this background would be reduced by a factor of about 150 compared
to KamLAND and would account to less than 10 events per year in the
LENA detector.

An additional background that imitates the geoneutrino signal is due
to $^9$Li, which is produced by cosmic muons in spallation reactions
with $^{12}$C and decays in a $\beta$-neutron cascade.  Only a small
part of the $^9$Li decays falls into the energy window which is
relevant for geoneutrinos. KamLAND estimates this background to be
$0.30 \pm 0.05$~\cite{kamland1}.  At CUPP the muon reaction rate would
be reduced by a factor $\simeq 10$ due to better shielding and this
background rate should be at the negligible level of $\simeq$~1 event
per year in LENA.

\section{Models of the Earth}
\label{sec:geomodels}

A detailed density profile of the Earth, the Preliminary Reference
Earth Model, was constructed by Dziewonski and Anderson in 1981 by
monitoring seismic activities~\cite{PREM}. Based on the examination
of meteorites and solar system materials (moon rocks and dust) and
available Earth material, geologists have deduced a model for the
distribution and concentration of elements in the Earth, the Bulk
Silicate Earth Model~\cite{macsun}. The most dominant and abundant
radioactive isotopes are $^{238}$U, $^{232}$Th and $^{40}$K; their
decays heat the Earth. The present-day total energy loss through the
Earth's surface is about 40~TW or 82~mW/m$^2$. The ratio of the
energy production due to radioactivity to the total heat flow at the
surface is known as the Urey ratio. In the Bulk Silicate Earth Model
this ratio is assumed to be 0.5, attributing 20~TW of the Earth's
heat loss to radioactivity. Other estimates take the Urey ratio to
be as large as~0.8~\cite{urey1, urey2}.

The flux of geoneutrinos is directly linked to the rate of radioactive
decays and to the generated heat. Therefore, it is of great interest
to measure the geoneutrino flux and thus deduce the main contributor
to the heat production. The elemental abundance ratios in the Bulk
Silicate Earth Model are Th/U${}\approx 4$ and K/U${}\approx 1.14
\times 10^{4}$. According to this model, radioactive isotopes are only
in the crust and mantle because they are lithophile whereas the core
is void of any significant amount of uranium, thorium or potassium.
Different estimates for their abundances in the crust and mantle
differ by factors of 2--3. An excellent overview of these abundances
and their spread is given in the GERM Reservoir Data Base~\cite{germ}.

An Italian group of physicists has, in cooperation with geologists,
constructed a reference model for the abundance values of uranium,
thorium and potassium. They used the values referenced in GERM and
derived a mean value for each element~\cite{mcfl}.  We will implement
these abundances into our Reference Model of the geoneutrino angular
distribution. For a discussion of the angular spectra of this model
and its uncertainties see Ref.~\cite{geonu}. Our Reference Model is in
accordance with the Bulk Silicate Earth constraint of a heat
production of 20~TW due to radioactive decays in the crust and mantle.

Other authors have speculated about the presence of uranium and
thorium in the core, for example in the context of a putative
georeactor~\cite{hern1,hern3}.  Moreover, the Earth's magnetic field
is not yet fully understood, but seems to be generated by a complex
interaction between core and mantle. The core itself is too hot to
sustain a permanent magnetic field. Therefore, it is assumed that
the magnetic field is powered by a geodynamo, where in a simplified
picture the magnetic field is generated by the motion of the liquid
outer core. One problem of this picture is the unknown energy source
for the geodynamo. Powering it by energy from the inner core leads
to cooling and solidification.  This process is constrained by the
present-day size of the inner core, implying that it must be younger
than 1.7~Gyr~\cite{french} much less than the Earth's age of
4.5~Gyr. Therefore, several authors have concluded that there are
radioactive elements in the core, providing heat and sustaining the
geodynamo~\cite{french}.

\begin{figure}
\begin{center}
\includegraphics[width=0.6\textwidth]{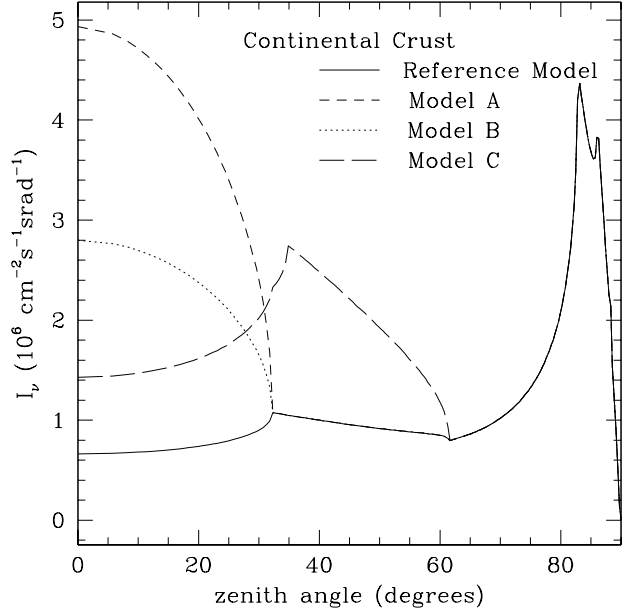}
\includegraphics[width=0.6\textwidth]{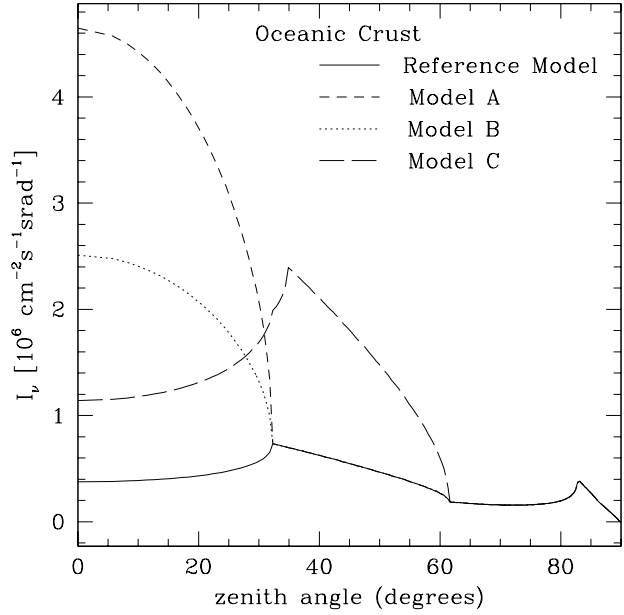}
\end{center}
\caption{Zenith-angle distribution of the geoneutrino flux where
$\theta=0$ corresponds to the vertical direction. Shown is the
total flux without flavor oscillations.
{\em Upper panel:}~Continental crust.
{\em Lower panel:}~Oceanic crust.}
\label{fig:2/3}
\end{figure}

The geoneutrino flux depends sensitively on location.  The oceanic
crust is depleted in radioactive elements, whereas the flux on the
continents is dominated by the crust. Thus an experiment situated in
the Pacific Ocean, e.g.~on Hawaii, would have better access to the
oceanic crust and the mantle.  For an experiment located on a
continent we have assumed a thickness of 50~km for the crust, implying
a total neutrino flux in our Reference Model of
$4.2\times10^6$~cm$^{-2}$~sec$^{-1}$ from uranium and
$4.1\times10^6$~cm$^{-2}$~sec$^{-1}$ from thorium decays.  For an
oceanic site we have chosen the crust to be rather thick (50~km), but
not included any sediments. If one wanted to determine the mantle
contribution, the oceanic crust would be a background to the
measurement so that the assumption of a thick oceanic crust is
conservative. The neutrino fluxes in this case are
$1.25\times10^6$~cm$^{-2}$~sec$^{-1}$ from uranium and
$0.88\times10^6$~cm $^{-2}$~sec$^{-1}$ from thorium decays.  The
zenith-angle distributions of the reference geoneutrino flux
corresponding to continental and oceanic locations are shown in
Fig.~\ref{fig:2/3}. Our assumption of a uniform oceanic crust with a thickness of 50 km has been made for computational reasons. Still, this model for the Hawaiian detector site is valid as our event rate is slightly lower than the rate found in \cite{mcfl}. In addition, as will become obvious in Sec.~\ref{sec:montecarlo}, changes in the crustal thickness of an oceanic site by even an order of magnitude can not be resolved by the detector.

Besides our reference model we consider three ``exotic'' cases A, B
and C, each of them either with a continental or an oceanic crust.
\begin{itemize}
\item[(A)] Fully radiogenic model with additional uranium and
thorium in the core, accounting for 20~TW additional heat production.
(Integrated neutrino flux increase of about 32\% relative to
the reference model in a continental location,
and 116\% in an oceanic location.)
\item[(B)] Same with 10~TW in the core. (Flux increase of
16\% and 58\%, respectively.)
\item[(C)] 20~TW in the Lower Mantle. (Flux increase of 41\% and
148\%, respectively.)
\end{itemize}
The zenith-angle distributions of the neutrino fluxes from these
models have been determined along the lines of Ref.~\cite{geonu} and
is shown in Fig.~\ref{fig:2/3}.

\begin{table}
\caption{Annual event rates for $2.5\times10^{33}$ target protons.
Flavor oscillations have been included with a global reduction
factor of~0.57.} \label{tab:fermi}
\medskip
\begin{ruledtabular}
\begin{tabular}{lll}
Model& Continental Crust& Oceanic Crust\\
\hline
Reference Model&        $1.02\times 10^3$ &  0.29$\times 10^3$\\
(A) 20 TW core&         1.35$\times 10^3$ & 0.62$\times 10^3$\\
(B) 10 TW core&         1.19$\times 10^3$ &  0.45$\times 10^3$\\
(C) 20 TW Lower Mantle& 1.44$\times 10^3$ & 0.71$\times 10^3$\\
\end{tabular}
\end{ruledtabular}
\end{table}

To obtain the event rate in a scintillator detector, neutrino flavor
oscillations have to be accounted for by including a global
$\bar\nu_e$ survival-probability factor of~0.57 as measured by
KamLAND~\cite{kamland1}. Matter effects for oscillations are not
important because of the small geoneutrino energies. Moreover, for
geoneutrino energies of 1.8--3.2~MeV and $\Delta
m^2=7.9\times10^{-5}~{\rm eV}^2$ the vacuum oscillation length is
57--101~km. Including distance and energy dependent survival
probabilities is a negligible correction to a global reduction
factor~\cite{fermi}. The annual event rates corresponding to our
models, including the reduction factor, are shown in
Tab.~\ref{tab:fermi} for a 50 kton detector with a fiducial volume
corresponding to $2.5\times 10^{33}$ protons.

Up to now we have assumed that the exotic heat source in the Earth's
core is caused by uranium and thorium decays, i.e.~the neutrino
spectrum from this additional source was taken to be identical with
the geoneutrino spectrum from the crust and mantle. However, the
possibility of a natural reactor in the Earth's core
(``georeactor'') has been discussed in the
literature~\cite{hern1,hern3}. In this case the neutrino flux could
be similar to that from an ordinary power reactor with energies
reaching up to about 8~MeV.  With this assumption the total $4
\pi$-georeactor neutrino flux can be estimated to be $\Phi_\nu
\simeq 1.9 \times 10^{23}~{\rm s}^{-1}$ for a thermal power of 1~TW.
Taking into account neutrino oscillations, the distance to the
center of the Earth, and the detection cross section we calculate an
event rate of about $210~{\rm y}^{-1}~{\rm TW}^{-1}$ in LENA. At
Pyh\"asalmi one would observe about 2200 events per year due to
neutrinos from nuclear power plants. Assuming a systematic uncertainty for the neutrino flux from the power plants of 6.5\%, as suggested in~\cite{kamland1}, we conclude that LENA will be able to identify a georeactor of $\ge 2 \,{\rm TW}$ after one year of measurement with a $3\sigma$ significance. The influence of the uncertainty of the mixing angle $\theta_{12}$ is negligible, because the flux of both the georeactor and the power plants depends on $\theta_{12}$ in the same way. 

\section{Monte-Carlo Study}
\label{sec:montecarlo}

To study the power of directional discrimination of a large
liquid-scintillator detector we have performed a Monte-Carlo
simulation of a large number of geoneutrino events and the
corresponding directional reconstruction.  We have assumed that the
detector response is independent of the event location, i.e.\ only the
spatial separation between the event $\bar\nu_e+p\to n+e^+$ and the
location of neutron capture is relevant. However, as pointed out in
chapter II, we consider a position resolution of point like events
located at the central axis of the detector.  As the light yield and
hence also the position resolution increases for off-axis events our
assumption is conservative. We have assumed that, on average, the
neutron capture point is displaced by 1.9~cm in the forward direction
relative to the $e^+$ event in agreement with the CHOOZ
measurement~\cite{chooz19}. Moreover, we have assumed that neutron
diffusion before capture causes a Gaussian distribution around this
mean value with a width $\sigma_x=\sigma_y=\sigma_z=4.0$~cm for an
unloaded PXE-based scintillator, whereas 2.4~cm is taken for a
Gd-loaded scintillator as described in Sec.~\ref{sec:LENA}.

\begin{figure}
\begin{center}
\includegraphics[width=0.5\textwidth]{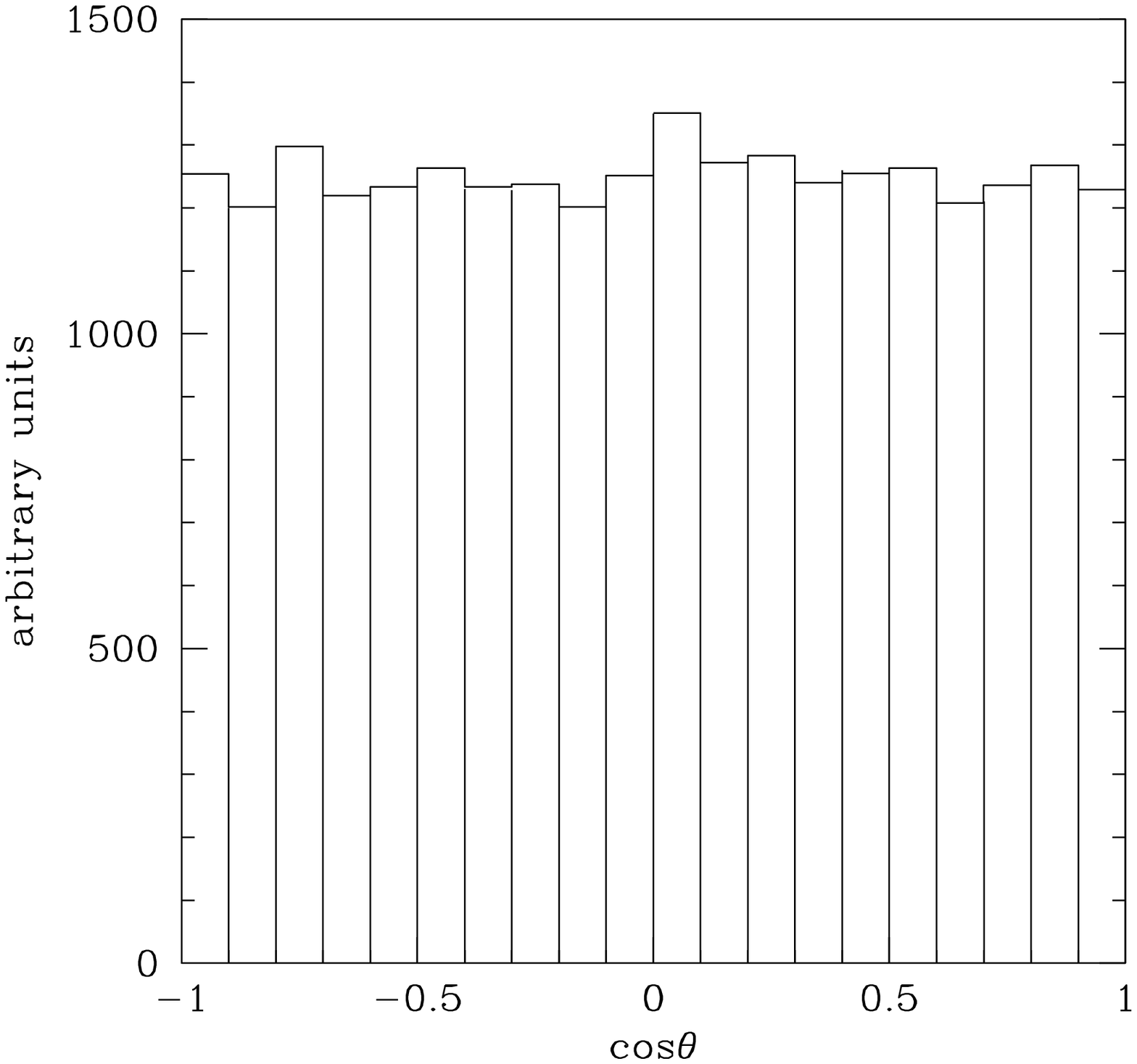}
\includegraphics[width=0.5\textwidth]{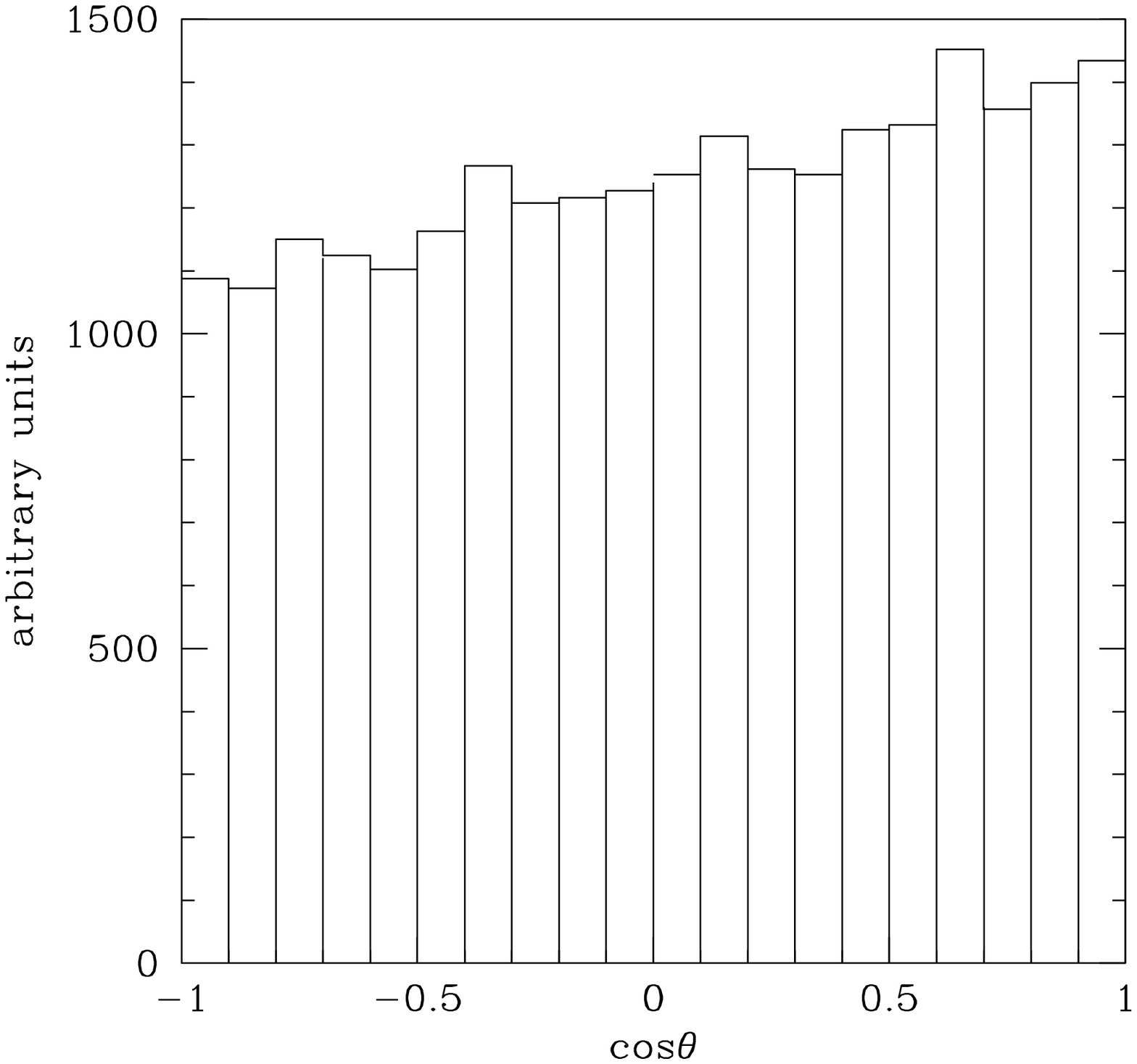}
\caption{\label{fig:4/5}Monte-Carlo example of reconstructed
zenith-angle distribution of 25,000 events in a Gd-loaded detector.
{\em Upper panel:} Neutrinos injected from the horizontal direction
($\cos\theta=0$). {\em Lower panel:} Neutrinos injected from the
vertical direction ($\cos\theta=1$).}
\end{center}
\end{figure}

In addition to this distribution, the main uncertainty originates from
the reconstruction of both events. For the positron event we have
assumed that the reconstructed location follows a Gaussian
distribution with a width given by Eq.~(\ref{eq:sigma_positron}).  The
actual spread of relevant visible energies is small so that we have
always used $E_{\rm vis}=1.4$~MeV as a typical value.  For an unloaded
scintillator, the reconstruction of the neutron event introduces an
even larger uncertainty; we have used a distribution as in
Fig.~\ref{fig:1} appropriate for the given light yield.  For a
Gd-loaded scintillator the uncertainty can be well fitted by a
Gaussian distribution with a width given by
Eq.~(\ref{eq:sigma_positron}).

For the sake of illustration we discuss the reconstructed zenith-angle
distributions for the two extreme cases where all neutrinos come from
the horizontal direction or all of them come vertically from below,
i.e\ with $\cos\theta=1$.  Figure~\ref{fig:4/5} displays the
reconstructed zenith-angle distribution of both cases, each generated
with 25,000 neutrino events and using a Gd-loaded detector with a
light yield $N_{\rm pe}=300$~pe/MeV. We conclude that, given the
relatively poor angular reconstruction capability of scintillator
detectors, the only angular-distribution information that can be
extracted is the slope of the distributions shown in
Fig.~\ref{fig:4/5}. Put another way, one can extract the total event
rate and the dipole contribution of the angular distribution, whereas
a determination of higher multipoles is unrealistic.  Therefore, we
write the reconstructed zenith-angle distribution in the form
\begin{equation}
\frac{d\dot N}{d\cos\theta}=\dot
N\,\left(\frac{1}{2}+p\,\cos\theta\right)
\end{equation}
where the event rate $\dot N$ and the coefficient $p$ are the two
numbers that characterize a given configuration of geophysical model
and detector type.

The event rates for our fiducial detector size with $2.5 \times
10^{33}$ target protons and different geophysical models have
already been reported in Tab.~\ref{tab:fermi}. What remains to be
determined by means of a Monte-Carlo simulation are the
corresponding coefficients $p$ and their uncertainty.  In
Tab.~\ref{tab:dipole} we show the results for $p$ for different
cases, always assuming a light yield of 300~pe/MeV. The uncertainty
$\sigma_p$ of the measured $p$ value scales with the inverse square
root of the number of events $N$ so that $s_p=\sigma_p\sqrt{N}$ is a
quantity independent of $N$. The value of $s_p$ can be derived
analytically for $p=0$, yielding
\begin{equation} \label{eq:s}
s_p=\frac{\sqrt{3}}{2}=0.866,
\end{equation}
which is valid for all $p\ll1$. We have checked with our Monte Carlo
that Eq.~(\ref{eq:s}) indeed applies to all $p$ values of interest
to us.

\begin{table}
\caption{Coefficient $p$ for the reconstructed zenith-angle
  distribution for different Earth models and different detector
  types, always assuming a light yield of
  300~pe/MeV.}\label{tab:dipole}
\medskip
\begin{ruledtabular}
\begin{tabular}{lll}
Model&\multicolumn{2}{l}{Coefficient $p$ for scintillator detectors}\\
&Unloaded PXE&Gd-loaded\\
\hline
{\bf Continenal Crust}\\
\quad Reference Model& 0.0283& 0.0377    \\
\quad (A) 20 TW core& 0.0377&  0.0521    \\
\quad (B) 10 TW core& 0.0333&  0.0459     \\
\quad (C) 20 TW Lower Mantle& 0.0351&0.0485\\
{\bf Oceanic Crust}\\
\quad Reference Model&0.0468& 0.0646      \\
\quad (A) 20 TW core& 0.0597& 0.0824      \\
\quad (B) 10 TW core& 0.0560& 0.0772     \\
\end{tabular}
\end{ruledtabular}
\end{table}

The number of events it takes to distinguish at the $1\sigma$ level
between an isotropic event distribution ($p=0$) and the actual
coefficient is given by $N_{1\sigma}=(s_p/p)^2=(3/4)\,p^{-2}$. For our
reference model at a continental site we find $N_{1\sigma}\approx 500$
events, for an oceanic site about 200 events. In order to distinguish
a geophysical model $i$ from model $j$ at the $1\sigma$ level, the
required number of events is
\begin{equation}
N_{1\sigma}=\frac{2s_{p}^2}{(p_i-p_j)^2}=\frac{3}{2}\,
\frac{1}{(p_i-p_j)^2}.
\end{equation}
A detection at the $n\sigma$ level requires $n^2$ times more events.

\begin{figure}
\begin{center}
\includegraphics[width=0.6\textwidth]{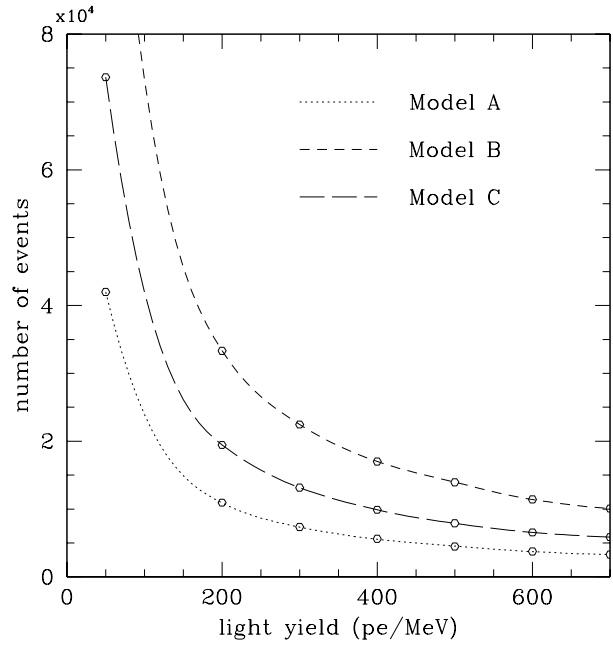}
\includegraphics[width=0.6\textwidth]{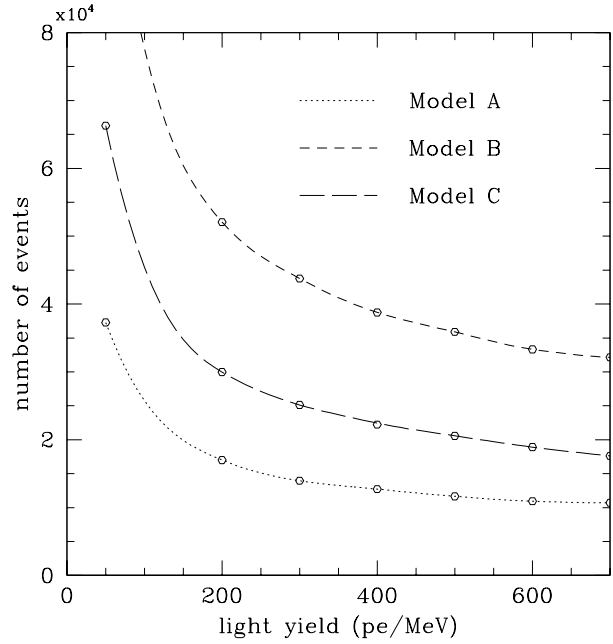}
\end{center}
\caption{Number of events needed to distinguish between models A, B or
C and the continental-crust reference model at $1\sigma$
significance. The points correspond to the values calculated with the
Monte Carlo. {\em Upper panel:} Gd-loaded scintillator. {\em Lower
panel:} Unloaded PXE-type scintillator.} \label{fig:6/7}
\end{figure}

\begin{figure}
\begin{center}
\includegraphics[width=0.6\textwidth]{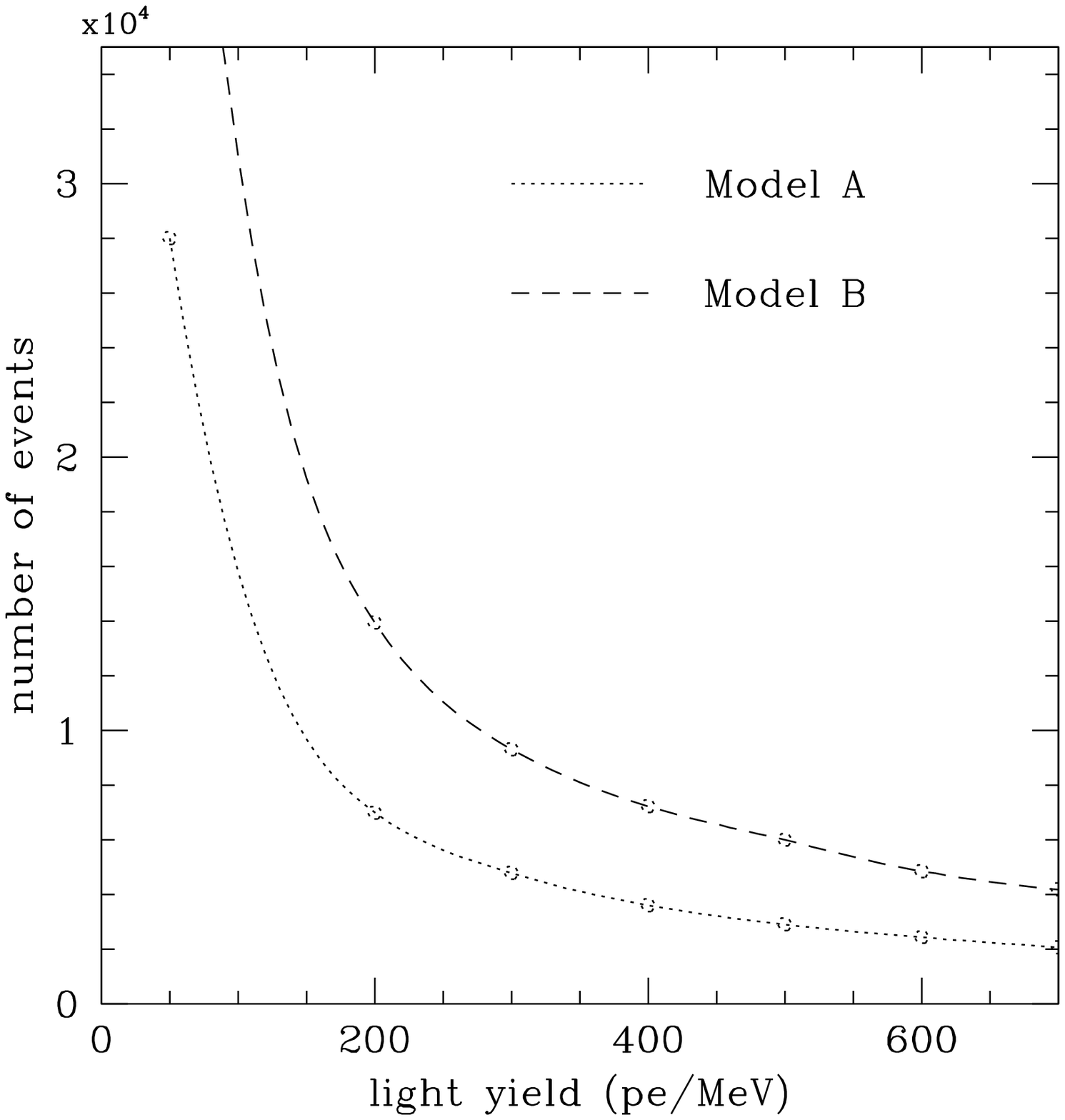}
\includegraphics[width=0.6\textwidth]{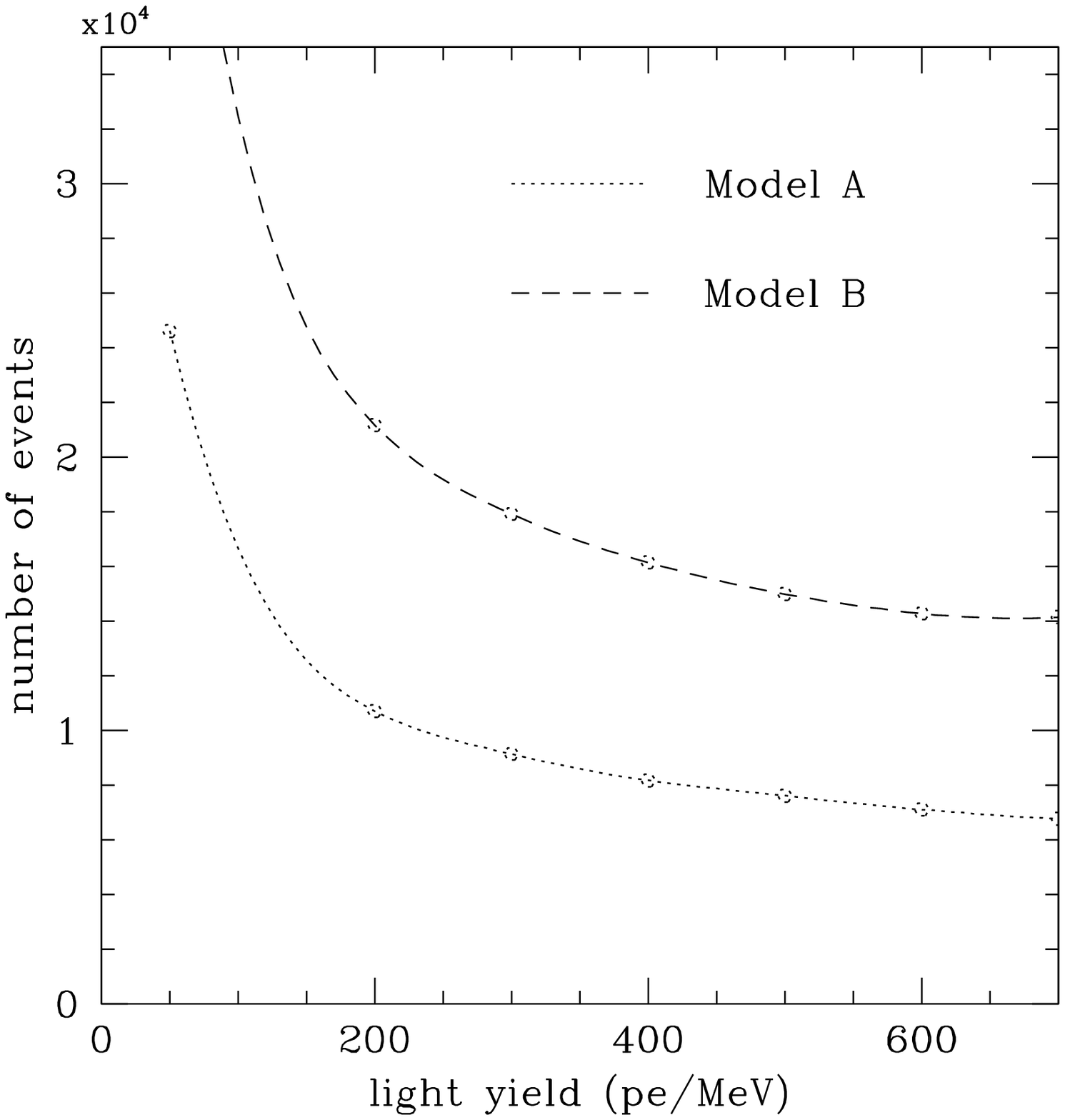}
\end{center}
\caption{Same as Fig.~\ref{fig:6/7} for oceanic crust. Model~C was
not included because the neutrino flux has almost the same angular
distribution as the reference model so that almost infinitely many
events are needed for a discrimination.} \label{fig:8/9}
\end{figure}

In the same way as for Tab.~\ref{tab:dipole} we have calculated the
slope $p$ for different light yields of the scintillator and have
determined the number of events it takes to distinguish each of the
exotic models from the reference case. In Fig.~\ref{fig:6/7} we
display $N_{1\sigma}$ for these cases and the continental-crust
situation as a function of the light yield $N_{\rm pe}$, both for an
unloaded PXE-type detector and a Gd-loaded one. In Fig.~\ref{fig:8/9}
we show the same for the oceanic crust. In the oceanic location we do
not show model~C because it corresponds to an increased flux from the
mantle, i.e.\ it is essentially identical with the reference model
except for the total flux and thus can not be distinguished on the
basis of the zenith-angle distribution.

Of course, the time required to achieve this discriminating power
depends on the detector size. For our fiducial volume with $2.5
\times 10^{33}$ target protons as in LENA one needs to scale with
the event rates shown in Tab.~\ref{tab:fermi}. In a
continental-crust location, all models produce an event rate of
roughly 1000 events per year, in full agreement with the KamLAND
measurement~\cite{kamland1}. Assuming for example a realistic light yield of 120 pe/MeV for the LENA detector (unloaded) one would have to measure roughly 150 year to distinguish even our most optimistic model A from the reference model. Therefore, even with optimistic
assumptions a 50~kt detector would need several decades for
distinguishing in a meaningful way between different geophysical
models on the basis of the angular event distribution. Moreover,
detector backgrounds should be included in a realistic assessment.

\section{Conclusions}
\label{sec:conclusions}

A future large-volume scintillator detector such as the proposed 50~kt
LENA would provide a high-statistics measurement of the geoneutrino
flux. The event rate would depend strongly on the detector location,
notably on whether an oceanic site such as Hawaii is chosen where a
reference event rate of about 300 per year (50~kt scintillator) is
expected or a continental site such as the Pyh\"asalmi mine in Finland
where the reference rate would be about 1000 per year. Therefore, the
geoneutrino flux could be measured with high significance and would
allow one to distinguish between different geophysical models.

The forward displacement of the neutron in the inverse beta decay
detection reaction provides directional information on the geoneutrino
flux. We have studied if this effect can be used to distinguish
between different geophysical models, notably if one could diagnose a
strong exotic energy source in the Earth's core under the assumption
that its neutrino spectrum is identical with that emitted by the crust
and mantle. While a deviation from an isotropic flux can be
ascertained with high significance, we find that a 50~kt detector is
too small to distinguish between different geophysical models on the
basis of the directional information alone, except perhaps for extreme
cases and optimistic assumptions about the detector performance.

In our study we have only used the neutrino flux from the Earth,
ignoring the contribution from power reactors because it depends
strongly on location. For example, in Pyh\"asalmi the neutrino flux
from power reactors adds roughly 25\% to the counting rate in the
energy window relevant for geoneutrinos. This contribution is not
negligible, but it does not change our overall conclusions.

We have also estimated the sensitivity of a LENA type detector for
determining a hypothetical georeactor in the Earth's core. As a
possible location the CUPP underground laboratory in Pyh\"asalmi
(Finland) was chosen and the background due to nuclear power plants
was calculated. At CUPP a 2~TW georeactor could be identified at a
statistical level of 3$\sigma$ after only one year of
measurement.

In summary, large-volume scintillator detectors of the next generation
will be extremely useful to study the interior of the Earth in the
``light of neutrinos''. However, the prime information will be the
total geoneutrino flux and its spectrum. It would be extremely
challenging to use the directional information alone to distinguish
between different geophysical models.

\begin{acknowledgments}
  We thank E.~Lisi for crucial discussions of an earlier version of
  this paper.  Partial support by the Maier-Leibnitz-Laboratorium
  (Garching), the Virtual Institute for Dark Matter and Neutrinos
  (VIDMAN, HGF), the Deutsche Forschungsgemeinschaft under Grant
  No.~SFB-375 and the European Union under the ILIAS project, contract
  No.~RII3-CT-2004-506222, is acknowledged.
\end{acknowledgments}



\begin{thebibliography}{99}

\bibitem{Araki:2005qa}
  T.~Araki {\it et al.},
  ``Experimental investigation of geologically produced
  antineutrinos with KamLAND,''
  Nature {\bf 436} (2005) 499.

\bibitem{Fiorentini:2005mr}
  G.~Fiorentini, M.~Lissia, F.~Mantovani and R.~Vannucci,
  ``Geo-neutrinos: A new probe of earth's interior,''
  Earth Planet.\ Sci.\ Lett.\  {\bf 238} (2005) 235
  [arXiv:physics/0508019].


\bibitem{Fiorentini:2005cu}
  G.~Fiorentini, M.~Lissia, F.~Mantovani and R.~Vannucci,
  ``How much uranium is in the Earth? Predictions for geo-neutrinos at
  KamLAND,''
  Phys.\ Rev.\ D {\bf 72} (2005) 033017
  [hep-ph/0501111].

\bibitem{geonu}
  B.~D.~Fields and K.~A.~Hochmuth,
  ``Imaging the Earth's interior: The angular distribution of
  terrestrial neutrinos,''
  hep-ph/0406001, accepted for publication in Earth, Moon and Planets

\bibitem{deMeijer:2004wq}
  R.~J.~de Meijer, E.~R.~van der Graaf and K.~P.~Jungmann,
  ``Quest for the nuclear georeactor,''
  Nucl.\ Phys.\ News {\bf 14} (2004) 20
  [physics/0404046].

\bibitem{Oberauer:2005kw}
  L.~Oberauer, F.~von Feilitzsch and W.~Potzel,
  ``A large liquid scintillator detector for low-energy
  neutrino astronomy,''
  Nucl.\ Phys.\ Proc.\ Suppl.\  {\bf 138} (2005) 108.

\bibitem{Undagoitia:12uu}
  T.~Marrod\'an Undagoitia {\it et al.},
  ``Search for the proton decay $p\to K^+\bar\nu$ in the
  large liquid scintillator low energy neutrino astronomy detector
  LENA,''
  Phys.\ Rev.\ D {\bf 72} (2005) 075014
  [hep-ph/0511230].

\bibitem{chooz}
  M.~Apollonio {\it et al.}  [CHOOZ Collaboration],
  ``Determination of neutrino incoming direction in the CHOOZ
  experiment  and its application to supernova explosion location
  by scintillator detectors,''
  Phys.\ Rev.\ D {\bf 61} (2000) 012001
  [hep-ex/9906011].

\bibitem{vogel}
  P.~Vogel and J.~F.~Beacom,
  ``Angular distribution of neutron inverse beta decay,
  $\bar\nu_e+p\to e^++n$,''
  Phys.\ Rev.\ D {\bf 60} (1999) 053003
  [hep-ph/9903554].

\bibitem{Back:2004zn}
  H.~O.~Back {\it et al.}  [Borexino Collaboration],
  ``Phenylxylylethane (PXE): A high-density, high-flashpoint organic
  liquid scintillator for applications in low-energy particle and
  astrophysics experiments'',
  arXiv:physics/0408032.

\bibitem{wurm}
  M.~Wurm,
  ``Untersuchungen zu den optischen Eigenschaften eines
  Fl\"ussigszintillators und zum Nachweis von Supernovae Relic
  Neutrinos mit LENA'', Diploma thesis, TU M\"unchen, Germany (2005).

\bibitem{conc}
  L.~Oberauer, C.~Grieb, F.~von Feilitzsch, I.~Manno,
  ``Production of light concentrators for BOREXINO and its
  Counting Test Facility'',
  Nucl. Instrum. Meth. A {\bf 530} (2004) 453.

\bibitem{ctf}
  G. Alimonti {\it et al.} [Borexino Collaboration],
  ``A large-scale low-background liquid scintillation detector:
  The Counting Test Facility at Gran Sasso'',
  Nucl. Instrum. Meth. A {\bf 406} (1998) 411.

\bibitem{bugey}
  J.~Hejwowski and A.~Szymanski,
  ``Lithium loaded liquid scintillator,''
  Review of Scientific Instruments {\bf 32} (1961) 1057--1058.

\bibitem{Ardellier:2004ui}
  F.~Ardellier {\it et al.},
  ``Letter of intent for double-CHOOZ: A search for the mixing angle
  theta(13),''
  hep-ex/0405032.

\bibitem{chooz19}
  M.~Apollonio {\it et al.},
  ``Search for neutrino oscillations on a long base-line at the CHOOZ
  nuclear power station,''
  Eur.\ Phys.\ J.\ C {\bf 27}(2003) 331
  [hep-ex/0301017].

\bibitem{kamland1}
  T.~Araki {\it et al.}  [KamLAND Collaboration],
  ``Measurement of neutrino oscillation with KamLAND:
  Evidence of spectral distortion,''
  Phys.\ Rev.\ Lett.\  {\bf 94} (2005) 081801
  [hep-ex/0406035].

\bibitem{PREM}
  A.~M.~Dziewonski and D.~L.~Anderson,
  ``Preliminary Reference Earth Model,''
  Phys.\ Earth Planet.\ Interiors {\bf 25} (1981) 297.
  For a tabulation see
  {\tt \verb+http://solid_Earth.ou.edu/prem.html+}

\bibitem{macsun}
  W.~F.~McDonough and S.-S.~Sun,
  ``The composition of the Earth,''
  Chem. Geol. {\bf 120} (1995) 223.

\bibitem{urey1}
  C.~Stein, in: {\em Global Earth Physics: A Handbook of Physical
  Constants, AGU Reference Shelf 1}, edited by T.~J.~Ahrens
  (American Geophysical Union, Washington, 1995), p.~144.

\bibitem{urey2}
  D.~McKenzie and F.~Richter,
  J. Geophys. Res. {\bf 86} (1981) 11667.

\bibitem{germ}
  Earth Reference Data and Models,
  {\tt http://earthref.org}

\bibitem{mcfl}
  F.~Mantovani, L.~Carmignani, G.~Fiorentini and M.~Lissia,
  ``Antineutrinos from the Earth: The reference model and its
  uncertainties,''
  Phys.\ Rev.\ D {\bf 69} (2004) 013001
  [hep-ph/0309013].

\bibitem{hern1}
  J.~M.~Herndon,
  J. Geomagn. Geoelectr. {\bf 45} (1993) 423.

\bibitem{hern3}
  J.~M.~Herndon,
  ``Nuclear georeactor origin of oceanic basalt
  He-3/He-4, evidence and implications,''
  PNAS, March 18 (2003), Vol. 100, No.~6, pp.~3047--3050

\bibitem{french}
  S.~Labrosse {\it et al.},
  Earth Planet. Sci. Lett. {\bf 190} (2001) 111.

\bibitem{fermi}
  G.~Fiorentini, T.~Lasserre, M.~Lissia, B.~Ricci and S.~Sch\"onert,
  ``KamLAND, terrestrial heat sources and neutrino oscillations,''
  Phys.\ Lett.\ B {\bf 558} (2003) 15
  [hep-ph/0301042].

\end{thebibliography}
\end{document}